\documentclass[12pt,preprint]{aastex}

\received{}
\accepted{}
\journalid{}{}
\articleid{}{}
\paperid{}
\cpright{AAS}{2010}
\ccc{}

\shorttitle{HeH$^{+}$Photodissociation}
\shortauthors{S. Miyake & al}

\begin{document}

\title{Rovibrationally-resolved Photodissociation of HeH$^{+}$}

\author{S. Miyake, C. D. Gay and P. C. Stancil}
\affil{Department of Physics and Astronomy and Center for Simulational
Physics,\\ The University of Georgia, Athens, GA 30602-2451}
\email{shinyam@-, stancil@physast.uga.edu, cgay1383@gmail.com}


\begin{abstract}
Accurate photodissociation cross sections have been obtained for the
A$~^{\mathrm{1}}\Sigma^{\mathrm{+}}\leftarrow
\mathrm{X}~^{\mathrm{1}}\Sigma^{\mathrm{+}}$ electronic transition of
HeH$^{\mathrm{+}}$ using ab initio potential curves and dipole transition
moments.
Partial cross sections have been evaluated for all rotational transitions
from the vibrational levels $v''=0-11$ and over the entire accessible
wavelength range $\lambda\lambda100-1129$. 
Assuming a Boltzmann distribution of the rovibrational levels of the
X$~^{\mathrm{1}}\Sigma^{\mathrm{+}}$ state, photodissociation cross sections are
presented for temperatures between $500$ and $12,000$ K.
A similar set of calculations was performed for the pure rovibrational
photodissociation in the X$~^{\mathrm{1}}\Sigma^{\mathrm{+}}$ electronic ground state,
but covering photon wavelengths into the far infrared.
Applications of the cross sections to the destruction
of HeH$^{\mathrm{+}}$ in the early Universe and in UV-irradiated environments
such as primordial halos and protoplanetary
disks are briefly discussed.

\end{abstract}

\keywords{molecular data --- molecular processes --- early Universe --- ISM: molecules --- 
photon-dominated regions }

\section{Introduction}
Photodissociation is an important mechanism for the destruction of
interstellar molecules in diffuse and translucent clouds, in photodissociation
regions, in circumstellar disks, in protoplanetary disks, and many other
environments with an intense radiation field.
HeH$^{\mathrm{+}}$ is believed to be one of the first molecules formed in the
early Universe \citep{2002JPhB...35R..57L}.
Its abundance is kept low due to photodissociation by the cosmic background
radiation (CBR) field.
\cite{2008A&A...490..521S}, however, have suggested that
HeH$^{\mathrm{+}}$ can efficiently scatter CBR
photons and may generate detectable fluctuations in the cosmic microwave background.
Within the Galaxy, \cite{1982ApJ...255..489R} discussed the
possibility of observing HeH$^{\mathrm{+}}$ in gaseous nebulae.
However, to date it has not been detected.
Searches in the planetary nebula NGC 7027 have placed upper limits on
its abundance through observations of its
vibrational \citep{1988ApJ...326..899M} and rotational
\citep{1997MNRAS.290L..71L,2001ApJ...562..515D} lines.
Intensive UV radiation from the central white dwarf may result in efficient
photodissociation, providing insight into the low abundance of this molecular
ion.

A number of HeH$^+$ photodissociation calculations have been performed over the
last three decades \citep{1978JPhB...11.3349S,flo79,1982ApJ...255..489R,
bas84,dum09,sod09}. However, they were primarily focused on the A$~^1\Sigma^+ \leftarrow$
X$~^1\Sigma^+$ electronic transition from the ground vibrational level. The
latter work of \citet{bas84}, \citet{dum09}, and \citet{sod09} also considered
transitions to higher electronic states, while \citet{bas84} obtained cross
sections from vibrational levels $v''=0-5$, but for a single
rotational level.  As a consequence, the available HeH$^+$ cross section
dataset is far from complete, which hinders the ability to accurately model
its abundance. In all but the lowest density environments (e.g., diffuse
interstellar clouds), excited rovibrational levels of HeH$^+$ are expected to be significantly
populated.

To partially address the lack of a comprehensive dataset of HeH$^+$ cross
sections, we  have performed extensive photodissociation calculations for
both the X$~^1\Sigma^+ \leftarrow$ X$~^1\Sigma^+$,
\begin{equation}
{\rm HeH}^ +(v'',J'') + h\nu \rightarrow {\rm He}(1s^2~^1S) + {\rm H}^+,
\end{equation}
 and 
A$~^1\Sigma^+ \leftarrow$ X$~^1\Sigma^+$, 
\begin{equation}
{\rm HeH}^ +(v'',J'') + h\nu \rightarrow {\rm He}^+(1s~^2S) + {\rm H}(1s~^2S),
\end{equation}
transitions
using the most accurate available molecular data.
Calculations were performed for the full range of 162 bound rovibrational levels
($v'',J''$) in the ground electronic state using
quantum-mechanical techniques.
Local thermodynamic equilibrium (LTE)
cross sections, which assume a Boltzmann distribution of rovibrational
levels, are also computed.
In Section 2, an overview of the theory of molecular photodissociation is
presented as well as the adopted molecular data.
The partial cross sections as well as the LTE
photodissociation cross sections for HeH$^{\mathrm{+}}$ are presented in Section 3,
while Section 4 briefly highlights the astrophysical importance of the results.
Atomic units are used throughout unless stated otherwise.


\section{Theory and Calculation}

\subsection{Potential curves and the Dipole Transition Moments}
Accurate ab initio calculations of the ground electronic
X$~^{\mathrm{1}}\Sigma^{\mathrm{+}}$ potential energy including adiabatic
corrections have been made by \citet{1979JMoSp..75..462B} and are adopted here.
For internuclear distances $R>10$~a$_0$, a smooth fit to the ab initio
potentials have been performed using the average long-range interaction
potential,
\begin{equation}
V_{\mathrm{L}} = -\frac{\alpha_d}{2R^{\mathrm{4}}},
\end{equation}
where $\alpha_d$ is the static dipole polarizability of the neutral hydrogen
or helium atom. 
We adopt the values $\alpha_{\mathrm{H}}=4.5$ and $\alpha_{\mathrm{He}}=1.38309$.
For $R<0.9$ a$_{\mathrm{0}}$, the potential curves have been fit to the
short-range interaction potential form $A\exp(-BR)+C$.
For X$~^1\Sigma^+$, we obtain a dissociation energy from $v''=0, J''=0$ of
1.844~eV using a reduced mass of 1467.3243 a.u. This is very close to the accurate non-Born-Oppenheimer, nonrelativisitic computation
of 1.845~eV, which included first-order relativistic corrections, of \citet{stanke06}
and in agreement with the value of 1.844~eV obtained by \citet{1979JMoSp..75..462B}.

Ab initio calculations of the excited electronic
A$~^{\mathrm{1}}\Sigma^{\mathrm{+}}$ state are obtained from
\citet{1995CPL...236..177K} and smoothly fit by the same approach.
The A$~^{\mathrm{1}}\Sigma^{\mathrm{+}}$ potential energy is
shifted so that the asymptotic energies as $R\rightarrow \infty$ exactly
match the difference in experimental H and He ionization potentials.
As a consequence the maximum possible wavelength for a bound-free
A$~^1\Sigma^+ \leftarrow$ X$~^1\Sigma^+$
transition from the highest-lying rovibrational level in the ground electronic
state is less than 1129~\AA.

The electric dipole moment function for the ground state X$~^1\Sigma^+$ was obtained
from the calculations of \citet{jur95} over the range $0.9 <R < 30$ a$_0$, while the
A$~^1\Sigma^+ \leftarrow$ X$~^1\Sigma^+$ electric dipole transition moment function was
adopted from
\citet{1995CPL...236..177K} over the range of $1.0<R<30.0$ $a_{\mathrm{0}}$. At short-range,
both dipole moments were fit to the form $AR^2 + BR$. For $R>30$ a$_0$, the ground state
dipole function was fit to the form $4R/5$,
while the transition dipole moment was found to fit well to the form $3.7\times 10^4/R^7$.

\subsection{The photodissociation cross section}
The expression for the cross section for a bound-free transition from initial
state $i$ to final state $f$ for an electric dipole transition
is given by
\citep[e.g.,][]{rau02}
\begin{equation}
\sigma^{fi}(E_{\rm ph})=\frac{2\pi^2 e^{\mathrm{2}} \hbar}{mc}\frac{df}{dE_{\rm ph}},
\label{sig1}
\end{equation}
where $E_{\rm ph}$ is the photon energy and $m$ the mass of the electron.
The continuum differential oscillator strength is defined in the length gauge as
\begin{equation}
\frac{df}{dE_{\rm ph}}=\frac{2m}{3\hbar^2} E_{\rm ph} |\langle \Phi_f(\vec{r},\vec{R}) |\vec{r} |
\Phi_i(\vec{r},\vec{R} \rangle|^2,
\label{dfde}
\end{equation}
where $\Phi(\vec{r},\vec{R})$ is the total molecular wave function, $\vec{r}$ the electronic
coordinate vector, and $\vec{R}$ the internuclear vector. Inserting Eq.~(\ref{dfde})
into Eq.~(\ref{sig1}) and using the definition of the fine-structure constant,
$\alpha=e^2/\hbar c$, gives
\begin{equation}
\sigma^{fi}(E_{\rm ph})=\frac{4\pi^2}{3}\alpha E_{\rm ph}  |\langle \Phi_f(\vec{r},\vec{R}) |\vec{r} |
\Phi_i(\vec{r},\vec{R} \rangle|^2.
\label{sig2}
\end{equation}
Taking all quantities in Eq.~(\ref{sig2}) in atomic units, but expressing
the cross section in cm$^2$, the numerical value of the pre-factor
becomes $2.689\times 10^{-18}$. Applying separation of variables for the
electronic and nuclear coordinates of $\Phi$,
the electric dipole transition moment function is defined as
\begin{equation}
D^{fi}(R)=\langle \phi_f(\vec{r}|R) | \vec{r} |\phi_i(\vec{r}|R) \rangle,
\label{dipole}
\end{equation}
where integration is taken over all electronic coordinates and $\phi(\vec{r}|R)$
is the electronic molecular wave function for fixed $R$.
The rotational photodissociation cross section from initial rovibrational
level $v''J''$ can then be written as
\begin{equation}
\sigma^{fi}_{v''J''}(E_{\rm ph})=2.689 \times 10^{\mathrm{-18}}
E_{\rm ph}  ~~\sum_{J'}  \left ( \frac{1}{2J''+1}
S_{J'}|D^{fi}_{k'J',v''J''}|^{\mathrm{2}} \right )
\mathrm{cm}^{\mathrm{2}},
\end{equation}
\citep[e.g.,][]{1988kirby},
where the H\"{o}nl-London factors, $S_{J'}(J'')$, can be expressed for
a ${\Sigma \leftarrow \Sigma}$ 
electronic transition as 
\begin{eqnarray}
S_{J'}(J'')&=&
\left\{
\begin{array}{ll}
J''-1,&~J'=J''-1~(\mbox{P-branch})\\
J''+1,&~J'=J''+1~(\mbox{R-branch}).
\end{array} \right .   
\end{eqnarray}
$D^{fi}_{k'J',v''J''}=
\langle \chi_{k'J'}(R)|D^{fi}(R)|\chi_{v''J''}(R)\rangle$ is the matrix element of the electric
dipole transition moment for absorption from the rovibrational level $v''J''$ in state
$i$ to the continuum $k'J'$ in state $f$,
with the integration taken over $R$.
$J$ is the angular momentum of nuclear motion, and 
$g$ is the degeneracy factor
given by
\begin{equation}
g=\frac{2-\delta_{\mathrm{0,\Lambda'+\Lambda''}}}{2-\delta_{\mathrm{0,\Lambda''}}},
\end{equation}
where $\Lambda'$ and $\Lambda''$ are the angular momenta projected along the
nuclear axis for the final and initial electronic states, respectively.
The continuum wave functions $\chi_{k'J'}(R)$ are normalized such that
they behave asymptotically as
\begin{equation}
\chi_{k'J'}(R) \sim  \sin(k'R-\frac{\pi}{2}J'+\eta_{J'})
\end{equation}
where $\eta_{J'}$ is the phase shift.
The bound and continuum rovibrational wave
functions, $\chi_{v''J''}(R)$ and $\chi_{k'J'}(R)$, respectively,
are solutions of the radial Schr$\ddot{{\rm o}}$dinger equation for nuclear
motion on the $i$ and  $f$ state potential curves, respectively.
The wave functions were obtained numerically using the standard Numerov method
\citep{1961MComput..15..363,1967JCoPh...1..382B,1977JChPh..67.4086J} 
with a step size of 0.001 a$_0$ over internuclear distances
$0.1<R<200$~a$_{\mathrm{0}}$.

\section{Results and discussion}

\subsection{Partial cross sections for the A$\leftarrow$X electronic photodissociation}
\begin{figure}
\plotone{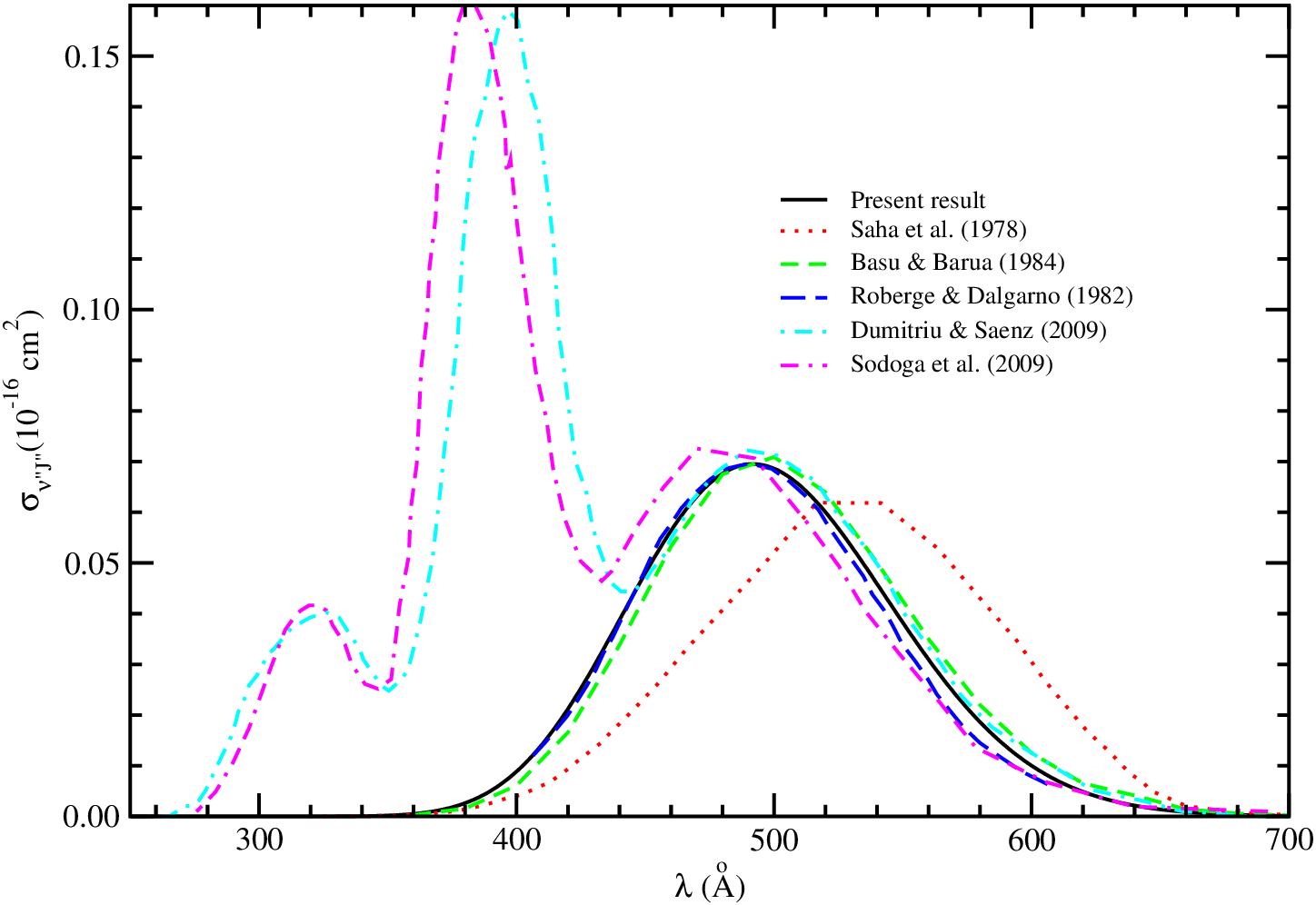}
\caption{The partial photodissociation cross sections $\sigma_{v''J''}$ as a
function of wavelength $\lambda$ for the A$~^1\Sigma^{\mathrm{+}}\leftarrow
\mathrm{X}~^1\Sigma^{\mathrm{+}}$ transition of HeH$^{\mathrm{+}}$ from the
vibrational level $v''=0$ (with $J''=1$). Solid line, current calculation; 
dotted line, \citet{1978JPhB...11.3349S}; dashed line \citet{bas84};
long dashed line, \citet{1982ApJ...255..489R}; dot-dash line, \citet{dum09}; dash-dot-dash
line, \citet{sod09}. Note the latter two calculations also include
contributions from higher excited electronic states which give
contributions for $\lambda \lesssim 450$~\AA.
\label{fig1}}
\end{figure}

\begin{figure}
\plotone{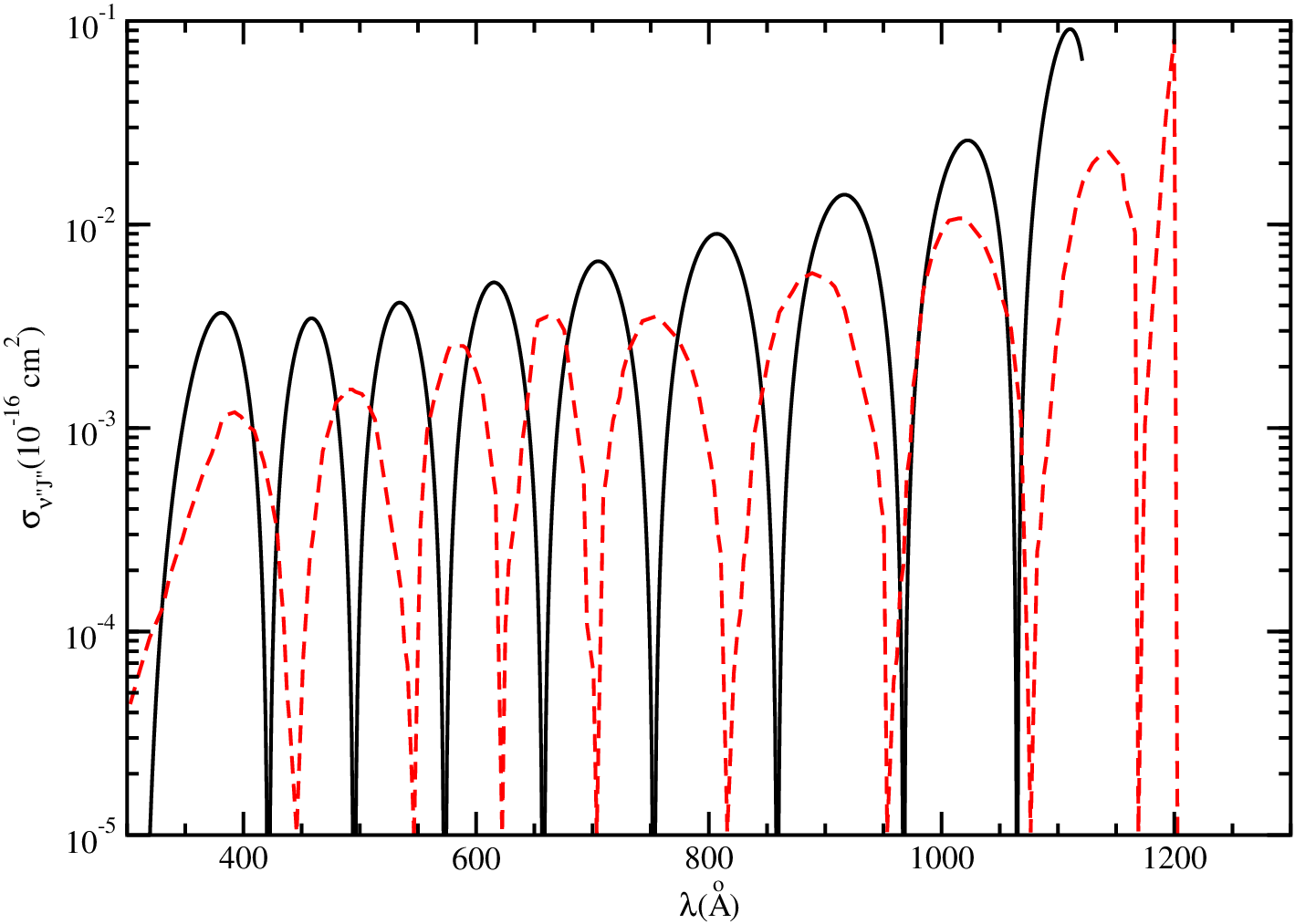}
\caption{Same as Fig.1, but for $v''=8$ ,$J''=1$. Solid line, current calculation; 
dashed line, \citet{1978JPhB...11.3349S}.
\label{fig2}}
\end{figure}

\begin{figure}
\plotone{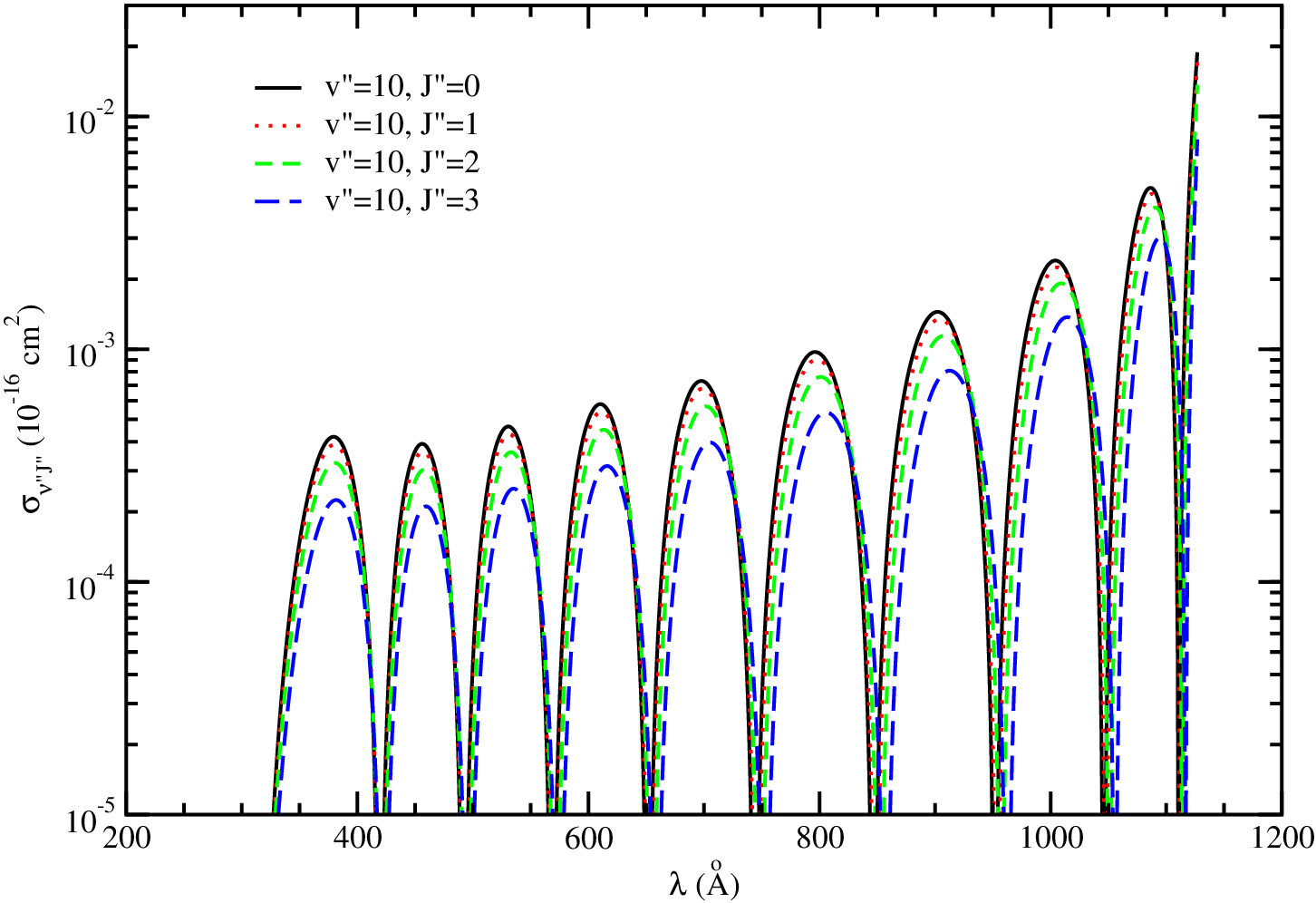}
\caption{The partial photodissociation cross sections $\sigma_{v''J''}$ for the
A$~^1\Sigma^{\mathrm{+}}\leftarrow \mathrm{X}~^1\Sigma^{\mathrm{+}}$
transition of HeH$^{\mathrm{+}}$ for $v''=10$ and all $J''$.
\label{fig3}}
\end{figure}

\begin{figure}
\plotone{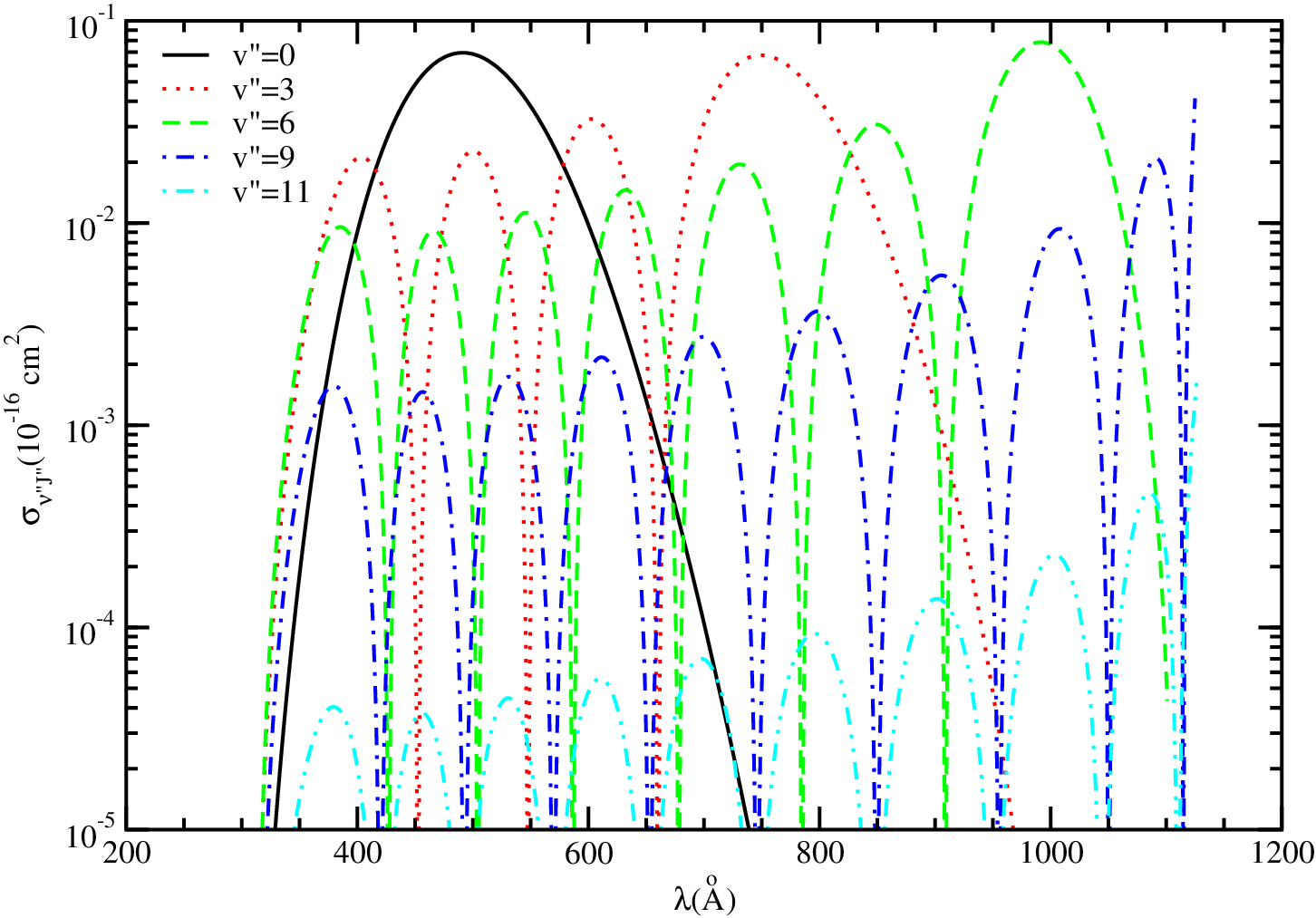}
\caption{Same as Fig. 3, but for $J''=0$ and select $v''$.
\label{fig4}}
\end{figure}

A sampling of the partial cross sections $\sigma_{v''J''}$ for the
A~$^{\mathrm{1}}\Sigma^{\mathrm{+}}\leftarrow
\mathrm{X}~^{\mathrm{1}}\Sigma^{\mathrm{+}}$ transition are presented in Figures
\ref{fig1}-\ref{fig4} as a function of the photon wavelength $\lambda$.
In Figure \ref{fig1}, the partial cross sections for photodissociation of
HeH$^{\mathrm{+}}$ from the rovibrational level $v''=0, J''=1$ are presented and
compared to previous calculations. The current calculations are shown to be
in excellent agreement with the earlier results of \citet{1982ApJ...255..489R} and \citet{bas84},
while the cross section peak obtained by 
\citet{1978JPhB...11.3349S} is seen to be shifted to longer wavelengths.
The more recent calculations of \citet{dum09} and \citet{sod09} include
contributions from transitions to higher electronic states of HeH$^+$, which
are not considered in the current work. In most astrophysical environments,
there are few photons at such short wavelengths so that our neglect
of the higher electronic transitions is reasonable. Nevertheless, there
is excellent agreement between the current calculations and those of
\citet{dum09} for $\lambda> 440$~\AA\ and good agreement with
the results of \citet{sod09}, though their $A\leftarrow X$ peak is
shifted to shorter wavelengths.

Experimental studies of molecular photodissociation is largely an
unexplored area of research and we are aware of only one measurement
for HeH$^+$ performed at the free-electron laser in Hamburg by
\citet{ped07}. They obtained a cross section of $(1.4\pm 0.7)\times 10^{-18}$
cm$^2$, but for fragmentation into H$^+$ and He($n\ge 2$) at 320~\AA.
While the \citet{dum09} and \citet{sod09} cross sections shown in Figure~\ref{fig1}
also include the H($n\ge 2$) + He$^+$ fragmentation channels, \citet{sod09}
find good agreement with the measurements when only the H$^+$ + He
channels are considered. 
Indirectly, this suggests that the current calculations at somewhat longer wavelengths
are consistent with experiment.

Figure \ref{fig2} shows our results for the partial cross sections
from the
vibrational level $v''=8, J''=1$ in comparison to the 
earlier calculations of
\citet{1978JPhB...11.3349S}. As for the $v''=0$, $J''=1$ case, the
\citet{1978JPhB...11.3349S} cross sections are shifted to longer
wavelengths.
Given the ionization potentials of H and He and the binding energy of the
$v''=8, J''=1$ level, the photodissociation threshold is 1123 \AA, which is
consistent with the current results. Further, the current $v''=8, J''=1$ binding
energy is in excellent agreement with the value obtained by \citet{1998ApJ...508..151Z}.

Additional cross sections are presented in Figure \ref{fig3} for $v''=10$ and
all bound $J''$ and in Figure \ref{fig4} for $v''=0-11$, for $J''=0$. While not
shown, good agreement is found with the $A\leftarrow X$, $v''=1-5,J''=1$
cross sections of \citet{bas84}. Good agreement is also found with the
results of \citet{sod09} for $v''=0,J''=10$ and $v''=1,J''=0$, though their
cross sections are again somewhat shifted to shorter wavelengths.

We note that our cross sections include results from rovibrational
levels which are very near the dissociation limit for which there is,
as yet, no experimental evidence in some cases. These are typically the last rotational
level in each vibrational manifold including $v'',J''$ = (0,23), (1,21), (2,20),
(3,18), (4,16), (5,14), (6,12), (7,10), (8,7), and (9,5), but also all $v''=10$
and 11 levels. All but $v'',J''=(0,23)$ were obtained by \citet{1998ApJ...508..151Z}
using a similar approach, while the non-Born-Oppenheimer method of
\citet{stanke06} predicts that $v'',J''=(11,0)$ is bound. However, transitions
involving the $v'',J''$ = (0,23),  (1,21), and (2,20) levels have been 
detected in a number of experiments \citep[e.g.,][]{liu97}.


\subsection{LTE cross sections for A$\leftarrow$X photodissociation}\label{LTE}
The total quantum-mechanical cross section for photodissociation
as a function of both temperature $T$ and wavelength in local thermodynamic
equilibrium (LTE) is given by \citep[e.g.,][]{arg74}
\begin{equation}
\sigma(\lambda,T)=\frac{\sum_{v''}\sum_{J''}g_{iv''J''}
\exp{[-(E_{\mathrm{g}}-E_{v''J''})/k_{\mathrm{b}}T]}
\sigma_{v''J''}(\lambda)}{Q_{\mathrm{HeH^{+}}}(T)},
\label{siglte}
\end{equation}
where a Boltzmann population distribution is assumed for the rovibrational
levels in the electronic ground state,  $g_{iv''J''}=2J''+1$ is the total vibrational-rotational
statistical weight, $E_{\mathrm{g}}$ is the binding energy of the lowest
rovibrational level, $k_{\mathrm{b}}$ is the Boltzmann constant, and
$Q_{\mathrm{HeH^{+}}}(T)$ is the partition function given by
\begin{equation}
Q_{\mathrm{HeH^{+}}}(T)=\sum_{v''}\sum_{J''}g_{iv''J''}
\exp{[-(E_{\mathrm{g}}-E_{v''J''})/k_{\mathrm{b}}T]}.
\label{part}
\end{equation}
In both Eqs.~(\ref{siglte}) and (\ref{part}), $E_{\rm g}$ and the binding
energies $E_{v'',J''}$ are taken as positive quantities. We obtain
partition functions in excellent agreement with those computed by
\citet{eng05} for $T$=1-4000~K, but our values are somewhat smaller
for higher temperatures.

Figure \ref{fig5} displays LTE cross sections for the 
A$~^{\mathrm{1}}\Sigma^{\mathrm{+}}\leftarrow
\mathrm{X}~^{\mathrm{1}}\Sigma^{\mathrm{+}}$ photodissociation transition as a function of
the wavelength for
temperatures between $500$ and $12,000$~K as given by Eq.~(\ref{siglte}).
The threshold wavelength at 1129~\AA\ corresponds to the asymptotic energy gap
between the A~$^{\mathrm{1}}\Sigma^{\mathrm{+}}$ and
X~$^{\mathrm{1}}\Sigma^{\mathrm{+}}$ electronic states. 
As the temperature increases, the cross section
for wavelengths between $\sim$800 and 1129 \AA\ increases dramatically, though
the peak cross section near 500 \AA\ decreases. In particular, for $T>3000$~K, the
cross section longward of the hydrogen Lyman limit becomes significant.


\subsection{The partial cross sections for X$\leftarrow$X rovibrational photodissociation}
For irradiated environments with photon wavelengths much greater than the Lyman limit,
photodissociation via pure rovibrational transitions may be dominant although the
cross sections are typically small.
For example, in Figure \ref{fig6}, the partial cross section for
rovibrational photodissociation in the X$~^{\mathrm{1}}\Sigma^{\mathrm{+}}$
electronic state for $v''=8,J''=1$ is presented and
comparison is made with the earlier calculations of
\cite{1978JPhB...11.3349S}. As for the A$\leftarrow $X case,
there is a significant shift in the threshold wavelength likely due to
an inaccurate binding energy calculation in the earlier work.
Additional examples of rovibrational photodissociation for the
X$~^{\mathrm{1}}\Sigma^{\mathrm{+}}\leftarrow
\mathrm{X}~^{\mathrm{1}}\Sigma^{\mathrm{+}}$ transition are given in Figures
\ref{fig7} and \ref{fig8}.
Figure \ref{fig7} displays photodissociation from the ground rovibrational level
($v''=0,J''=0$) which has a threshold of 6732~\AA\ and a peak cross section
magnitude that is $\sim$10$^7$ times smaller than for the
A$~^{\mathrm{1}}\Sigma^{\mathrm{+}}\leftarrow \mathrm{X}~^{\mathrm{1}}\Sigma^{\mathrm{+}}$ electronic transition.
Figure \ref{fig7} also displays cross sections from $v''=0$ and a selection of $J''$.
Orbiting resonances due to quasi-bound levels are evident for $J''\gtrsim5$
near thresholds.
Figure \ref{fig8} displays similar results, but for $J''=0$ and a selection of
$v''$. We are unaware of other previous rovibrational photodissociation
calculations for HeH$^+$.

\subsection{LTE cross sections for X$\leftarrow$X rovibrational photodissociation}
Assuming a Boltzmann population distribution of rovibrational levels in the
ground electronic state, LTE cross sections are computed for gas
temperatures between 500 and 5000~K.
Figure \ref{fig9} shows LTE cross sections for the 
X$~^{\mathrm{1}}\Sigma^{\mathrm{+}}\leftarrow
\mathrm{X}~^{\mathrm{1}}\Sigma^{\mathrm{+}}$ photodissociation of
HeH$^{\mathrm{+}}$ as a function of the wavelength.
For $T$ less than $\sim$1000~K, the LTE cross section peaks at a
wavelength corresponding to the $v''=0,J''=0$ threshold.
Photons in this wavelength range are primarily responsible for photodestruction
of HeH$^{+}$.
For higher temperatures, the population in highly excited $v'',J''$ levels
contributes substantially to the cross section increasing its value at large
wavelength.

\section{Astrophysical Applications}
 The possible presence of
HeH$^+$ in astrophysical environments was probably first discussed by \citet{dab78},
while early models of its abundance and emission lines in gaseous nebulae
were performed by \citet{bla78} and \citet{flo79}. Later,
\citet{1982ApJ...255..489R} gave a comprehensive discussion of the mechanisms for the formation
and destruction of HeH$^{+}$ in astrophysical plasmas. Over the subsequent three decades,
the role of HeH$^+$ has been considered in a variety of environments. \citet{1998ApJ...508..151Z}
and \citet{sta98} discussed the formation of HeH$^+$ in supernova ejecta, planetary
nebulae, high-$z$ Ly$\alpha$ clouds, and broad-line clouds.
 Searches for
its rotational and vibrational emissions features, focusing on planetary nebula NGC 7027,
have so far been unsuccessful. However, \citet{mil92} proposed a tentative identification
in the ejecta of supernova 1987A, but this has yet to be confirmed. Here we discuss
three environments in which HeH$^+$  is expected to have a significant abundance and
the possible impact of the current photodissociation cross sections including the
early Universe, population III objects, and protoplanetary disks. 
We do not provide
photodissociation rates as they are sensitive to the properties of the local
radiation field including its shape, intensity, and attenuation via the local dust and
gas. Instead, we provide cross sections to facilitate calculation of {\it local}
photo-rates.

\begin{figure}
\plotone{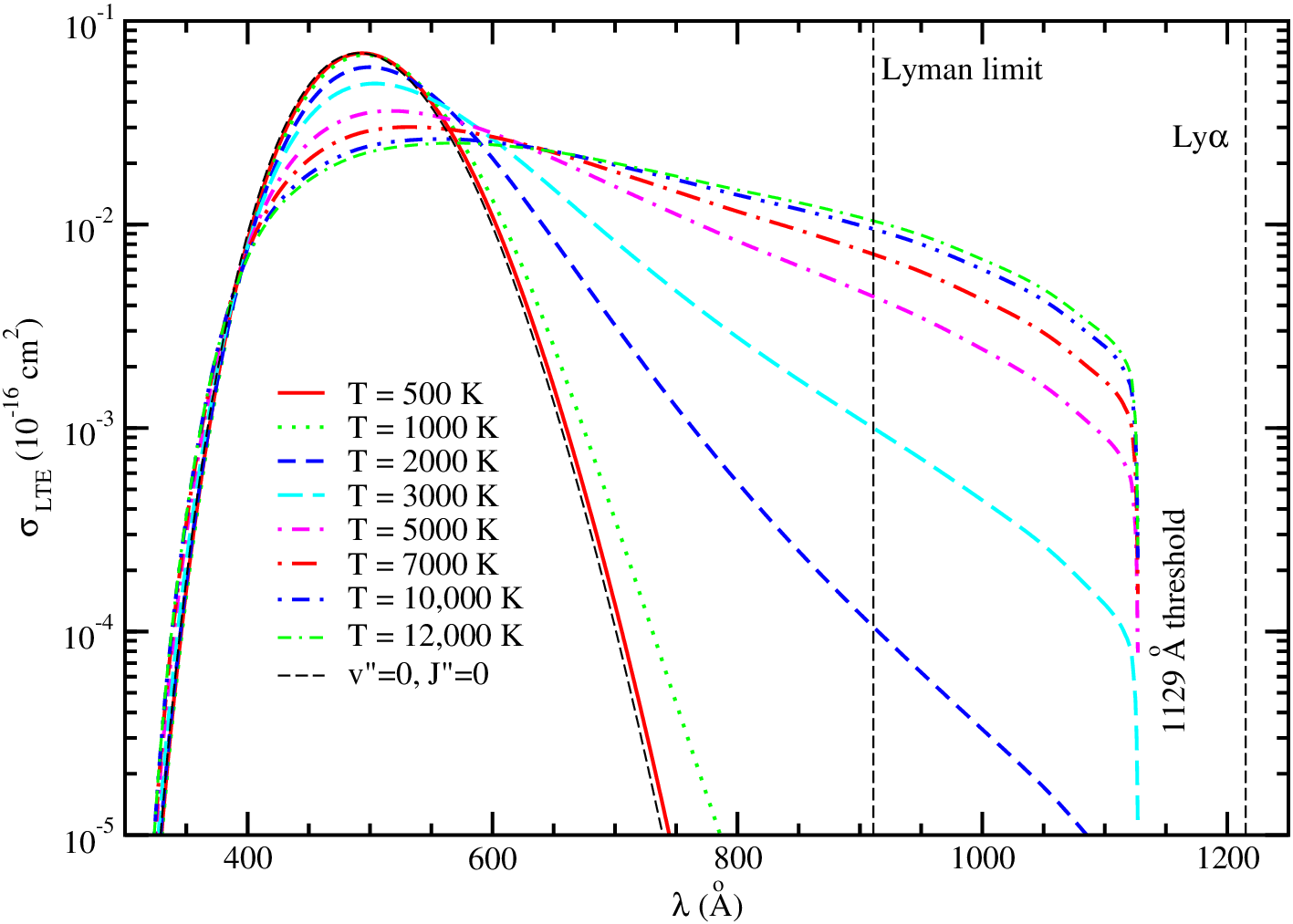}
\caption{Total HeH$^{+}$  A$~^1\Sigma^{\mathrm{+}}\leftarrow
\mathrm{X}~^1\Sigma^{\mathrm{+}}$ LTE photodissociation cross section for temperatures
from 500 to 12,000 K. The $v''=0$, $J''=0$ partial cross section is also plotted for comparison.
\label{fig5}}
\end{figure}

\begin{figure}
\plotone{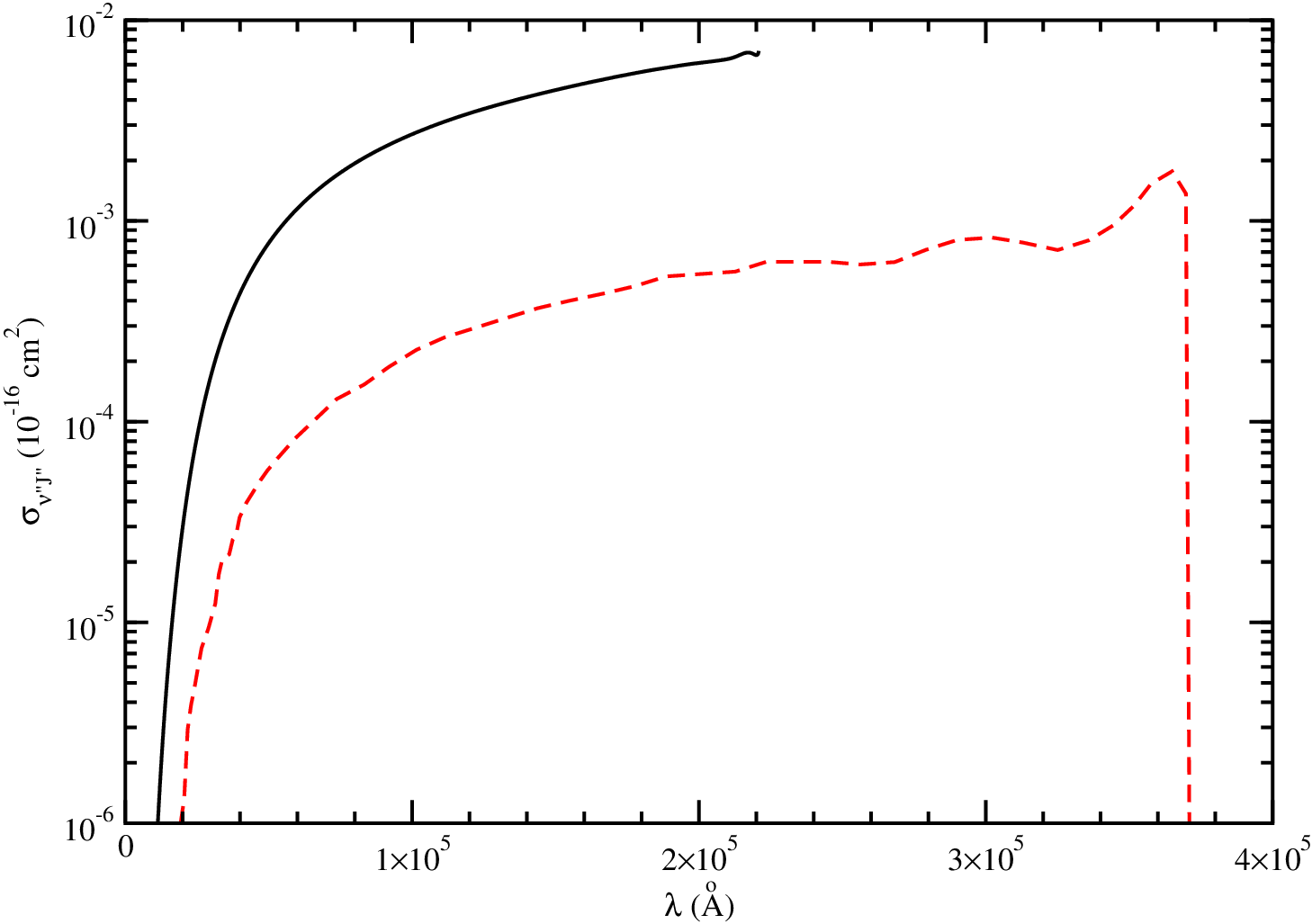}
\caption{The partial photodissociation cross section of $v''=8$, $J''=1$
for the X~$^1\Sigma^{\mathrm{+}}\leftarrow \mathrm{X}~^1\Sigma^{\mathrm{+}}$
rovibrational transition.  Solid line, current calculation;  dashed line,
\cite{1978JPhB...11.3349S}.
\label{fig6}}
\end{figure}

\begin{figure}
\plotone{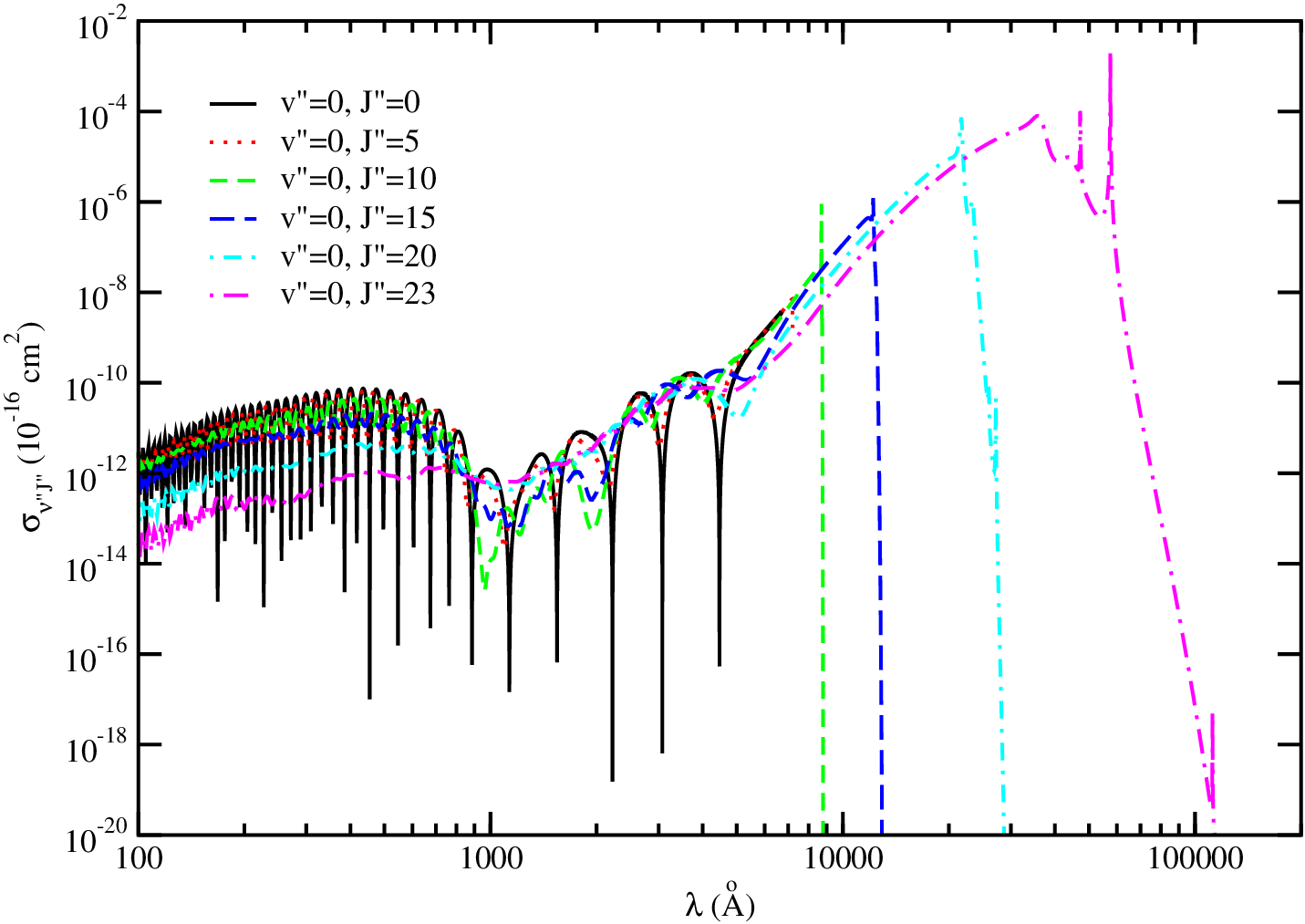}
\caption{The partial photodissociation cross sections
for the X~$^1\Sigma^{\mathrm{+}}\leftarrow \mathrm{X}~^1\Sigma^{\mathrm{+}}$
rovibrational transition for $v''=0$ and select $J''$. 
\label{fig7}}
\end{figure}

\begin{figure}
\plotone{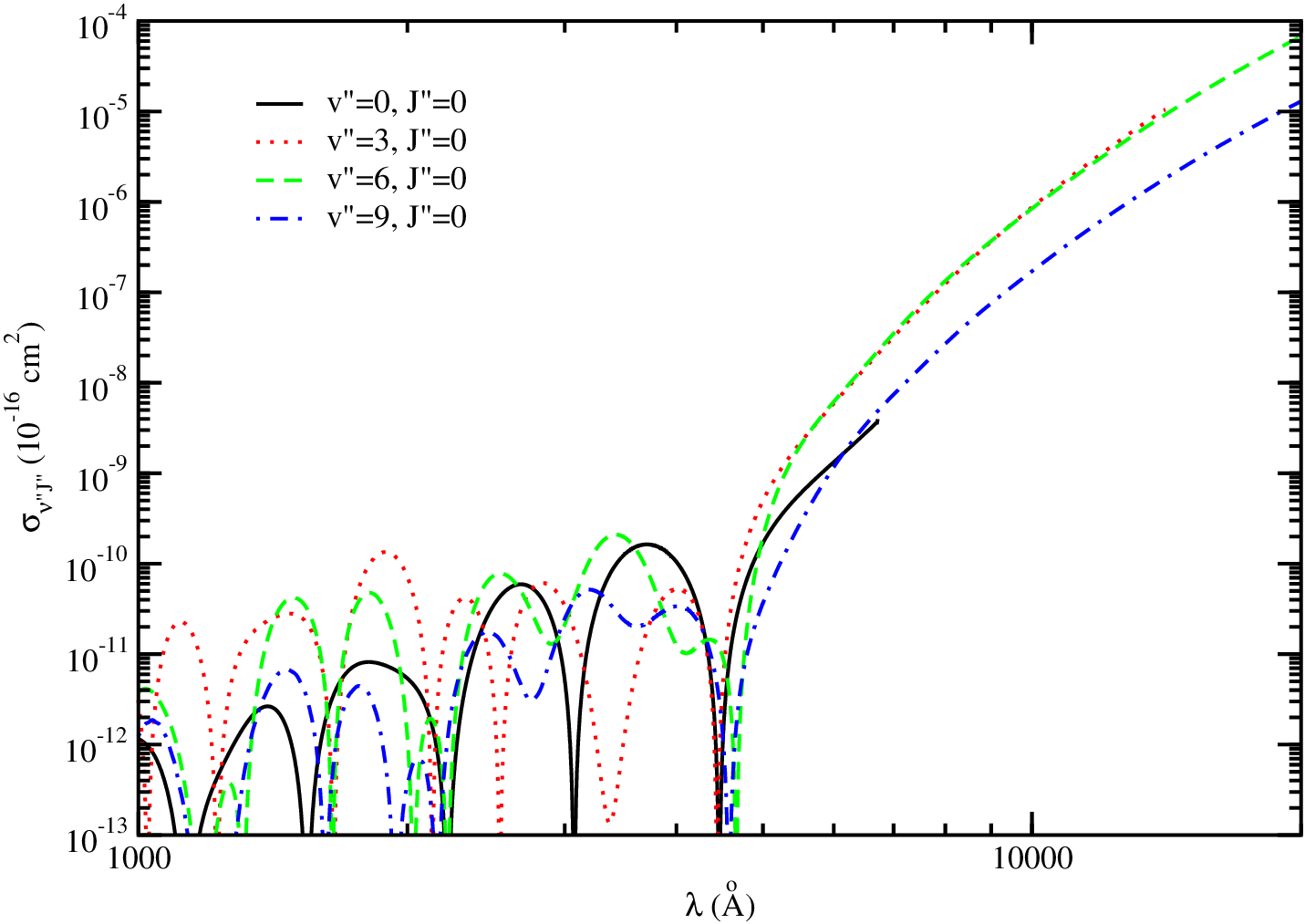}
\caption{Same as Fig. 7, but for $J''$=0 and select $v''$.
\label{fig8}}
\end{figure}

\begin{figure}
\plotone{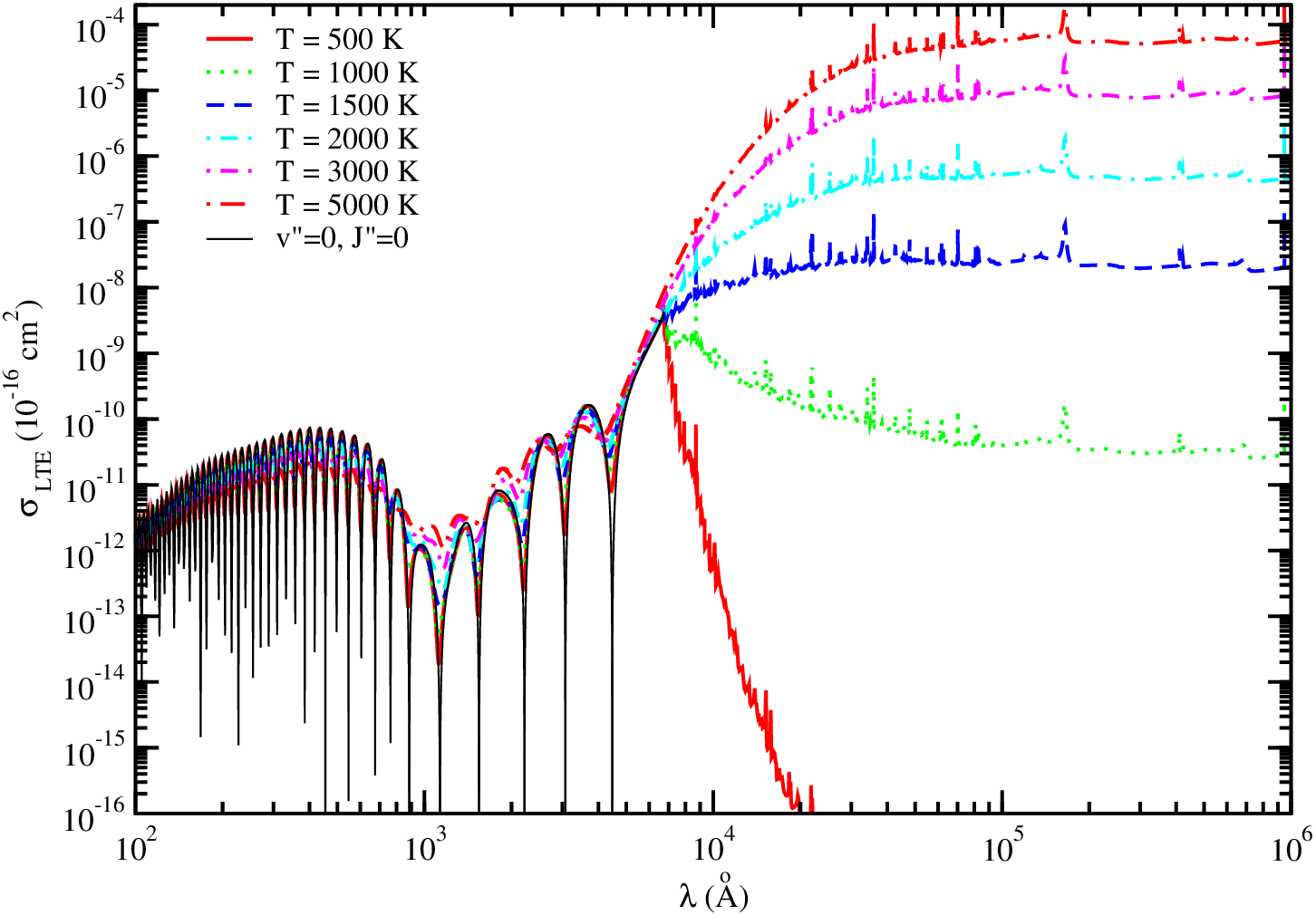}
\caption{Total HeH$^{+}$ X$~^1\Sigma^{\mathrm{+}}\leftarrow
\mathrm{X}~^1 \Sigma^{\mathrm{+}}$ LTE photodissociation cross section for temperatures
from 500 to 5000 K. The $v''=0$, $J''=0$ partial
cross section is also plotted for comparison.
\label{fig9}}
\end{figure}

\begin{figure}
\plotone{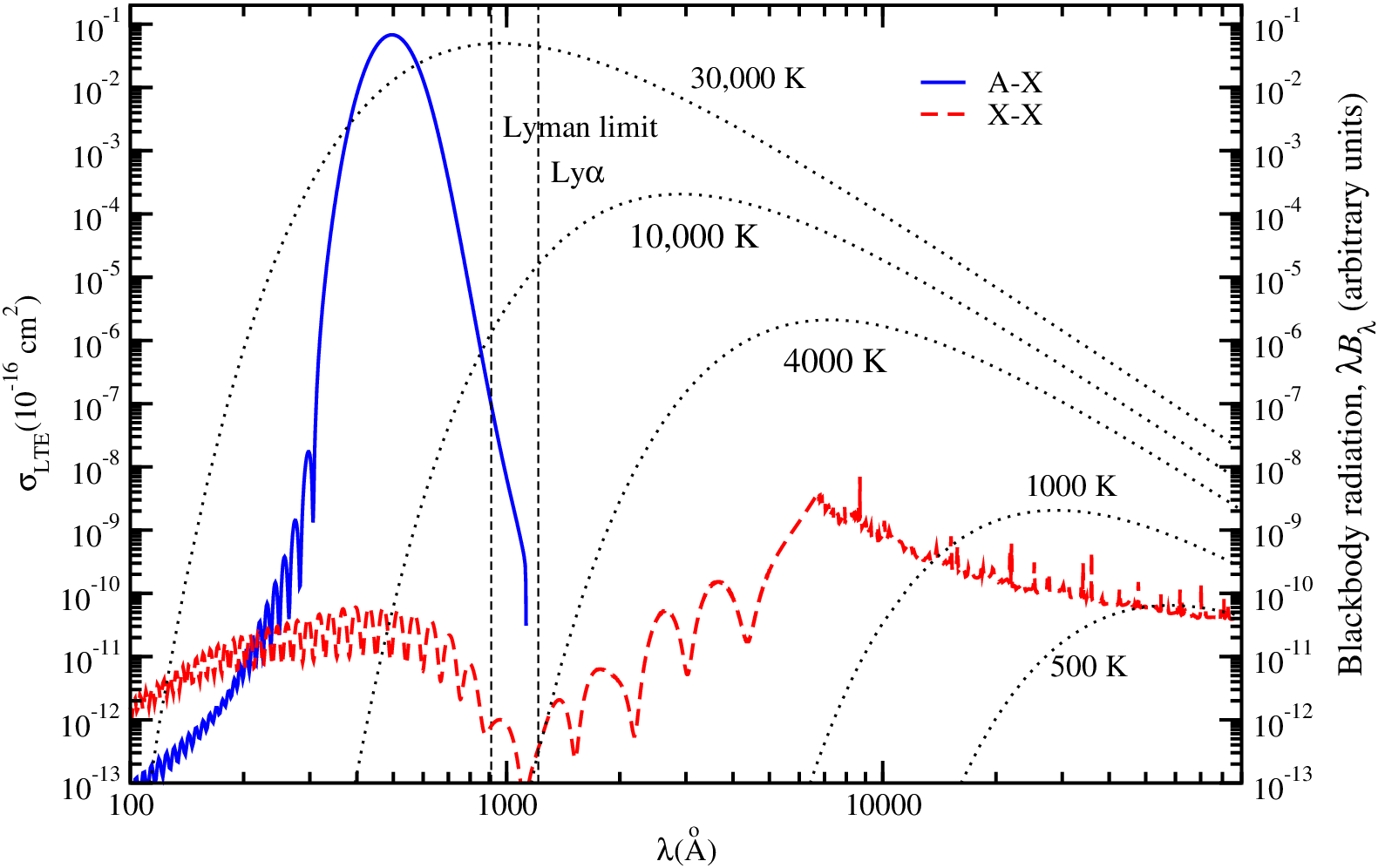}
\caption{Total HeH$^{+}$ LTE photodissociation cross section at 
1000 K for both the X~$^1\Sigma^{\mathrm{+}}\leftarrow \mathrm{X}~^1\Sigma^{\mathrm{+}}$
and A$~^1\Sigma^{\mathrm{+}}\leftarrow \mathrm{X}~^1\Sigma^{\mathrm{+}}$
transitions. Blackbody radiation curves (dotted lines) for various radiation temperatures
are plotted for comparison.
\label{fig10}}
\end{figure}

In the recombination era of the early Universe, radiation from the
CBR field  can
efficiently destroy HeH$^{+}$ through the
X$~^{\mathrm{1}}\Sigma^{\mathrm{+}}\leftarrow
\mathrm{X}~^{\mathrm{1}}\Sigma^{\mathrm{+}}$ rovibrational transition until a redshift of
$z\sim300$ \citep{gal98,1998ApJ...509....1S,2002JPhB...35R..57L,2008A&A...490..521S}.
Early Universe chemical models typically obtain the destruction rate via detail balance
from the radiative association rate coefficient. Given that HeH$^+$ has
a dipole moment, the CBR field will likely thermalize the rovibrational populations.
The current results, shown in Fig. \ref{fig9}, could be used to improve such
calculations by directly obtaining the photodissociation rate due to the CBR
field. For illustrative purposes, both the X$\leftarrow$X and A$\leftarrow$X LTE
cross sections for a gas temperature of 1000~K are plotted in Fig.~\ref{fig10} and
compared to blackbody radiation curves for a number of radiation temperatures $T_r$.
For $z$ between $\sim$1000 and 300 ($T_r\sim$3000 and 1000 K), the X$\leftarrow$X
transition dominates the photodissociation, but for higher $z$ the A$\leftarrow$X 
transition will become important (see also Fig. \ref{fig5}), 
though it is neglected in current models. For $1000 \lesssim z
\lesssim 3000$, models predict a plateau in the HeH$^+$ abundance
\citep{1998ApJ...508..151Z,1998ApJ...509....1S,2008A&A...490..521S}. While
the abundance at this redshift is small, it would likely be further suppressed
due to photodissociation via the A$\leftarrow$X transition.

In \citet{miy10}, we considered the photodetachment of H$^-$ due to
far UV (FUV) sources prior to the reionization epoch including
population III (Pop III) stars and the high-$z$ intergalactic medium (IGM)
radiation field. 
Once a Pop III star is formed, the HeH$^{+}$ present in its primordial
halo, which consists of baryonic and dark matter, will be exposed to an
intense FUV stellar radiation field.
As Pop III stars
are expected to be massive, they will emit essentially blackbody radiation with
effective temperatures $T_{\rm eff}$ between $\sim$30,000 and 10$^7$~K \citep{sch02}.
Figure \ref{fig10} illustrates that Pop III stellar radiation will photodissociate
HeH$^+$ through both the 
A$~^{\mathrm{1}}\Sigma^{\mathrm{+}}\leftarrow
\mathrm{X}~^{\mathrm{1}}\Sigma^{\mathrm{+}}$  and
X$~^{\mathrm{1}}\Sigma^{\mathrm{+}}\leftarrow
\mathrm{X}~^{\mathrm{1}}\Sigma^{\mathrm{+}}$ transitions with the A$\leftarrow$X
contribution increasing with $T_{\rm eff}$. \citet{1982ApJ...255..489R} computed
A$\leftarrow$X photodissociation rates for $20,000 \le T_{\rm eff} \le 500,000$ K
for complete blackbodies and with a 24.6 eV cut-off accounting for He I continuum
opacity. However, the Pop III FUV radiation will carve-out a surrounding H II region
resulting in a 13.6 eV cut-off at the Lyman limit (reducing the rate), but the
densities in the molecular region may be high enough for the HeH$^+$ level
populations to approach a thermal distribution. The latter effect is likely to
increase the rate, as suggested by Figs.~\ref{fig5} and \ref{fig10}, since  
\citet{1982ApJ...255..489R} only considered photodissociation from
$v''=0,J''=1$. Further, the  B$\leftarrow$X, C$\leftarrow$X, and higher
electronic transitions will play little role due to Lyman limit or He I cut-offs.
On the other hand, the high-redshift IGM radiation field is expected to have
a Lyman limit cut-off, to be modulated by H I resonant absorption, and to
behave as a power-law at longer wavelengths
\citep[e.g.,][]{miy10}, so that HeH$^+$ present in the outer molecular regions
of primordial halos exposed to the IGM field would be photodissociated primarily
through the X$~^{\mathrm{1}}\Sigma^{\mathrm{+}}\leftarrow
\mathrm{X}~^{\mathrm{1}}\Sigma^{\mathrm{+}}$ transition.

Photoprocesses are known to have a significant influence on
molecular gas in protoplanetary disks as they are exposed to
intense UV radiation from the young star \citep[e.g.,][]{van08} and the accretion disk
\citep{fra11} and to infrared radiation from dust grains \citep[e.g.,][]{gor04}.
As pointed out by \citet{van08}, the environment of a protoplanetary disk is
likely to be vastly different from average interstellar conditions including
i)  dust grains with different sizes and properties, ii) varying spectral shapes of
the incident UV radiation, iii)  higher intensity radiation, and other factors.
Therefore, pre-computed photo-rates cannot be readily applied. 
The gas temperature is typically $\sim$50~K in the mid-plane, but
increases with height and towards the star, reaching a few 1000 K
in the inner disk. The dust temperature decreases from $\sim$400~K in
the inner disk to $\sim$100 K in the mid-plane of the outer disk \citep{gor04}.
Effective stellar temperatures range from 4000~K for a T Tauri star to
$\sim$30,000~K \citep{van08}, while the accretion shock radiation
is expected to approach the Lyman limit with intensities $10^5$ to $10^7$
that of the standard interstellar radiation field \citep{fra11}. For the high
densities ($10^4$ to $10^7$ cm$^{-3}$) in the disk, the HeH$^+$ rovibrational
populations would be expected to be thermalized so that LTE photodissociation
cross sections would be appropriate. While \citet{van08} mention that HeH$^+$
cannot be photodissociated, Figure \ref{fig10} suggests that photodissociation
can occur due to dust and cooler stellar radiation through the X$\leftarrow$X
transition. Hotter stars, and in particular the intense accretion radiation,
may dominate the photodissociation rate through the A$\leftarrow$X transition
shortward of 1129~\AA. However, intense Ly$\alpha$,  from T Tauri stars,
which may contribute to the photodissociation of other molecular, will have a
negligible effect for HeH$^+$.

\section{Conclusions}
Using ab initio potentials and dipole moment functions,
accurate cross section calculations have been performed for
the photodissociation of HeH$^{\mathrm{+}}$ through the 
A$~^1\Sigma^{\mathrm{+}}\leftarrow
\mathrm{X}~^1\Sigma^{\mathrm{+}}$ and
X$~^1\Sigma^{\mathrm{+}}\leftarrow
\mathrm{X}~^1\Sigma^{\mathrm{+}}$ transitions.
The partial cross sections have been evaluated for all the rotational
transitions from the vibrational levels $v''=0-11$ of 
the X$~^1\Sigma^{\mathrm{+}}$ electronic state.
LTE cross sections are also calculated for  temperatures 
between $500$ and $12,000$~K.
The resulting cross sections are applicable
to the destruction of HeH$^{\mathrm{+}}$ in the early Universe and in UV
irradiated molecular regions including primordial halos, photodissociation
regions, and protoplanetary disks. To facilitate the calculation of local
photo-rates for particular astrophysical environments, all photodissociation
cross section data can be obtained from the UGA Molecular Opacity
Project website (http://www.physast.uga.edu/ugamop/).

\acknowledgments
This work was supported by NSF Grant AST-0607733 and 
HST-AR-11776.01-A  which was provided by NASA through
a grant from the Space Telescope Science Institute, which is operated
by the Association of Universities for Research in Astronomy, Incorporated,
under NASA contract NAS5-26555.




\clearpage

\clearpage

\clearpage

\clearpage

\clearpage

\clearpage

\clearpage

\clearpage

\clearpage

\clearpage

\begin{thebibliography}{}
%
\bibitem[Argyros(1974)]{arg74} Argyros, J. D. 1974, J. Phys. B, 7, 2025

\bibitem[Basu \& Barua(1984)]{bas84} Basu, D. \& Barua, A. K. 1984,
   J. Phys. B, 17, 1537
%

\bibitem[Bishop \& Cheung(1979)]{1979JMoSp..75..462B} Bishop, D.~M., \& 
Cheung, L.~M.\ 1979, J. Molec. Spectrosc., 75, 462 

\bibitem[Black(1978)]{bla78} Black, J. H. 1978, \apj, 222, 125
%
\bibitem[Blatt(1967)]{1967JCoPh...1..382B} Blatt, J.~M.\ 1967, J. Computat. Phys., 1, 382 
%
\bibitem[Cooley(1961)]{1961MComput..15..363} Cooley, J.~W.\ 1961, Math. Comput., 15, 363 
%

\bibitem[Dabrowski \& Herzberg(1978)]{dab78} Dabrowski, I., \& Herzberg, G. 1978,
    Trans. N. Y. Acad. Sci., 38, 14
\bibitem[Dinerstein \& Geballe(2001)]{2001ApJ...562..515D} Dinerstein, H.~L.,
\& Geballe, T.~R.\ 2001, \apj, 562, 515


\bibitem[Dumitriu \& Saenz(2009)]{dum09} Dumitriu, I. \& Saenz, A. 2009, J. Phys. B.
   42, 165101

\bibitem[Engel et al.(2005)]{eng05} Engel, E. A., Doss, N., Harris, G. J., \&
   Tennyson, J. 2005, \mnras, 357, 471

\bibitem[Flower \& Roueff(1979)]{flo79} Flower, D. R. \& Roueff, E. 1979, \aap, 72, 361

\bibitem[France, Yang, \& Linsky(2011)]{fra11} France, K., Yang, H., \& Linsky, J. L.
2011, \apj, 729, 7

\bibitem[Galli \& Palla(1998)]{gal98} Galli, D. \& Palla, F.  1998, \aap, 335, 403

\bibitem[Gorti \& Hollenbach(2004)]{gor04} Gorti, U. \& Hollenbach, D. 2004, \apj,
   613, 424

\bibitem[Johnson(1977)]{1977JChPh..67.4086J} Johnson, B.~R.\ 1977, \jcp, 
67, 4086 
\bibitem[Ju\v{r}ek, \v{S}pirko, \& Kraemer(1995)]{jur95} Ju\v{r}ek, M., \v{S}pirko, V, \& Kraemer, W. P.
   1995, Chem. Phys., 193, 287
\bibitem[Kirby \& van Dishoeck(1988)]{1988kirby}
Kirby, K. ~P. \& van Dishoeck, E. ~F. \ 1988, Adv. At. Mol. Phys., 25, 437 
%
\bibitem[Kraemer et al.(1995)]{1995CPL...236..177K} Kraemer, W.~P., 
\v{S}pirko, V., \& Ju\v{r}ek, M.\ 1995, Chemi. Phys. Lett., 236, 177 
%
\bibitem[Lepp et al.(2002)]{2002JPhB...35R..57L} Lepp, S., Stancil, P.~C., \&
Dalgarno, A.\ 2002, J. Phys. B Atomi., Molec., Phys., 35, 57
%
\bibitem[Liu et al.(1997)]{1997MNRAS.290L..71L} Liu, X.-W., et al.\ 1997,
\mnras, 290, L71

\bibitem[Liu \& Davies(1997)]{liu97} Liu, Z. \& Davies, P. B. 1997, \prl, 79, 2779

\bibitem[Miller et al.(1992)]{mil92} Miller, S, Tennyson, J., Lepp, S., \& Dalgarno,
   A. 1992, Nature, 335, 420

\bibitem[Miyake et al.(2010)]{miy10} Miyake, S., Stancil, P.  C., Sadeghpour, H. R.,
   Dalgarno, A., McLaughlin, B. M., \& Forrey, R. C. 2010, \apj, 709, L168

\bibitem[Moorhead et al.(1988)]{1988ApJ...326..899M} Moorhead, J.~M., Lowe,
R.~P., Wehlau, W.~H., Maillard, J.-P., \& Bernath, P.~F.\ 1988, \apj, 326, 899
\bibitem[Pedersen et al.(2007)]{ped07} Pedersen, H. B., et al. 2007, \prl, 98,
    223202 

\bibitem[Rau(2002)]{rau02} Rau, A. R. P. 2002, {\it Astronomy-inspired Atomic
   and Molecular Physics},
   Astrophys. Space Sci. Lib., Vol. 271, (Kluwer Acad. Publ., Dordrecht)

\bibitem[Roberge \& Dalgarno(1982)]{1982ApJ...255..489R} Roberge, W., \&
Dalgarno, A.\ 1982, \apj, 255, 489

\bibitem[Saha et al.(1978)]{1978JPhB...11.3349S} Saha, S., Datta, K.~K., 
\& Barua, A.~K.\ 1978, J. Phys. B Atomi., Molec., Phys., 11, 3349 

\bibitem[Schaerer(2002)]{sch02} Schaerer, D. 2002, \aap, 382, 28
\bibitem[Schleicher et al.(2008)]{2008A&A...490..521S} Schleicher, D.~R.~G.,
Galli, D., Palla, F., Camenzind, M., Klessen, R.~S., Bartelmann, M., \& Glover,
S.~C.~O.\ 2008, \aap, 490, 521
%

\bibitem[Sodoga et al.(2009)]{sod09} Sodoga, K., Loreau, J., Lauvergnat, D.,
   Justum, Y., Vacek, N., \& Desouter-Lecomte, M. 2009, \pra, 80, 033417

\bibitem[Stancil \& Dalgarno(1998)]{sta98} Stancil, P. C. \& Dalgarno, A.
   1998, Faraday Discuss., 109, 61

\bibitem[Stancil et al.(1998)]{1998ApJ...509....1S} Stancil, P.~C., Lepp, S.,
\& Dalgarno, A.\ 1998, \apj, 509, 1
%
\bibitem[Stanke et al.(2006)]{stanke06} Stanke, M., Kedziera, D., Molski, M,
   Bubin, S., Barysz, M., \& Adamowicz, L. 2006, \prl, 96, 233003

\bibitem[van Dishoeck, Jonkheid, \& van Hemert(2008)]{van08} van Dishoeck, E. F., 
   Jonkheid, B., \& van Hemert, M. C. 2008, arXiv:astro-ph/0806.0088

\bibitem[Zygelman et al.(1998)]{1998ApJ...508..151Z} Zygelman, B., Stancil,
P.~C., \& Dalgarno, A.\ 1998, \apj, 508, 151
%
\end{thebibliography}
\end{document}